\documentclass[aps,prl,twocolumn,amsmath,amssymb,superscriptaddress,nofootinbib,showpacs]{revtex4-1}
\usepackage{aas_macros}
\usepackage{graphicx}
\usepackage[colorlinks=true]{hyperref}
\newcommand\om{\omega_\textrm{max}}

\begin{document}

\title{Searching for an oscillating massive scalar field as a dark matter candidate using  atomic hyperfine frequency comparisons}

\author{A.~Hees}
\email{ahees@astro.ucla.edu}
\affiliation{SYRTE, Observatoire de Paris, PSL Research University, CNRS, Sorbonne Universit\'es, UPMC Univ. Paris 06, LNE, 61 avenue de l'Observatoire, 75014 Paris, France}
\affiliation{Department of Physics and Astronomy, University of California, Los Angeles, CA 90095, USA}

\author{J.~Gu\'ena}
\email{jocelyne.guena@obspm.fr}
\affiliation{SYRTE, Observatoire de Paris, PSL Research University, CNRS, Sorbonne Universit\'es, UPMC Univ. Paris 06, LNE, 61 avenue de l'Observatoire, 75014 Paris, France}

\author{M.~Abgrall}
\email{michel.abgrall@obspm.fr}
\affiliation{SYRTE, Observatoire de Paris, PSL Research University, CNRS, Sorbonne Universit\'es, UPMC Univ. Paris 06, LNE, 61 avenue de l'Observatoire, 75014 Paris, France}

\author{S.~Bize}\email{sebastien.bize@obspm.fr}
\affiliation{SYRTE, Observatoire de Paris, PSL Research University, CNRS, Sorbonne Universit\'es, UPMC Univ. Paris 06, LNE, 61 avenue de l'Observatoire, 75014 Paris, France}

\author{P.~Wolf}\email{peter.wolf@obspm.fr}
\affiliation{SYRTE, Observatoire de Paris, PSL Research University, CNRS, Sorbonne Universit\'es, UPMC Univ. Paris 06, LNE, 61 avenue de l'Observatoire, 75014 Paris, France}

\date{\today}

\pacs{04.50.Kd,04.80.Cc,06.20.Jr,95.35.+d}

\begin{abstract}
We use six years of accurate hyperfine frequency comparison data of the dual rubidium and caesium cold atom fountain FO2 at LNE-SYRTE  to search for a massive scalar dark matter candidate. Such a scalar field can induce harmonic variations of the fine structure constant, of the mass of fermions and of the quantum chromodynamic mass scale, which will directly impact the rubidium/caesium hyperfine transition frequency ratio. We find no signal consistent with a scalar dark matter candidate but provide improved constraints on the coupling of the putative scalar field to standard matter. Our limits are complementary to previous results that were only sensitive to the fine structure constant, and improve them by more than an order of magnitude when only a coupling to electromagnetism is assumed.
\end{abstract}

\maketitle

\par While thoroughly tested~\citep{will:1993fk,*will:2014la}, the theory of General Relativity (GR) is currently challenged by theoretical considerations and by galactic and cosmological observations. Indeed, the development of a quantum theory of gravitation or of a theory that would unify gravitation with the other fundamental interactions leads to deviations from GR. These modifications are usually characterized by the introduction of new fields in addition to the space-time metric to model the gravitational interaction. For example, string theory generically predicts the existence of new scalar fields (dilaton, moduli, axions). In addition, in the current cosmological paradigm, some galactic and cosmological observations are explained by the introduction of cold Dark Matter (DM) and of Dark Energy. Little is currently known about these two components that constitute the major part of our Universe. They can be interpreted as new types of matter (although they have not been directly detected so far), as a modification of the theory of gravitation or even as a combination of the two.

The introduction of nonminimally coupled scalar fields additionally to GR (tensor-scalar theories) generally leads to a space-time dependence of fundamental constants, which can then be searched for by experiments that test the Einstein equivalence principle (EEP) like weak equivalence principle (WEP) tests or tests of local position or Lorentz invariance (LPI and LLI) \cite{will:1993fk}. In the past, spectroscopy of different atomic transitions has been widely used to carry out such searches, and has set the tightest limits so far on a possible present-day space-time variation of fundamental constants \cite{van-tilburg:2015fj,ashby:2007dq,blatt:2008kx,cingoz:2007qf,fortier:2007bh,rosenband:2008fk,guena:2012ys,leefer:2013xy,godun:2014sf,huntemann:2014nr,ferrell:2007mz,blatt:2008kx,guena:2012ys,tobar:2013gf,peil:2013ul,leefer:2013xy}.

Such scalar fields could be a candidate for DM and/or dark energy. Different cosmological evolutions of the scalar fields are possible (see e.g. \cite{damour:1994uq, minazzoli:2014xz}). In several scenarios (in particular in the one defined by the action below), a massive scalar field will oscillate at a frequency related to its mass, leading to a corresponding oscillation of fundamental constants (see e.g. \cite{arvanitaki:2015qy,Stadnik2015a}). Recently atomic spectroscopy of Dy has been used to constrain such oscillations \cite{van-tilburg:2015fj} of the fine structure constant $\alpha$. In this letter we present limits on possible oscillations of a linear combination of constants ($\alpha$, quark mass and $\Lambda$ quantum chromodynamics -- QCD -- mass scale) using $\approx$ six years of highly accurate hyperfine frequency comparison of $^{87}$Rb and $^{133}$Cs atoms. This provides complementary constraints to those from Dy spectroscopy \cite{van-tilburg:2015fj} which is sensitive to $\alpha$ alone. When assuming a variation of $\alpha$ only, our results improve the limits of \cite{van-tilburg:2015fj} by over an order of magnitude. 

\par Tensor-scalar theories of gravitation have been widely studied as an extension of GR (see for example~\cite{jordan:1949vn,brans:1961fk,fujii:2003fi,damour:1992ys,damour:1996fk} and references therein) motivated by unification theories~\cite{green:1988oa,damour:1994fk,damour:1994uq,gasperini:2002kx,damour:2002vn,*damour:2002ys} or by models of Dark Energy~\cite{ratra:1988vn,caldwell:1998kx,peebles:1993fk,hees:2012kx}. Moreover, models of a light scalar field coupled to DM have been proposed~\cite{damour:1990fk,alimi:2008zr,carroll:2009fh,carroll:2010xw,mohapi:2015pi} as well as bosonic models of DM~\cite{preskill:1983fv,abbott:1983db,weinberg:1978rc}. In this letter, we focus on a massive scalar field model of DM parametrized by the action (see e.g. \cite{damour:2010zr})
\begin{align}
	S=&\frac{1}{c}\int d^4x \frac{\sqrt{-g}}{2\kappa}\left[R-2g^{\mu\nu}\partial_\mu\varphi\partial_\nu\varphi-V(\varphi)\right]\label{eq:action} \\
	&  +\frac{1}{c}\int d^4x \sqrt{-g} \left[ \mathcal L_\textrm{SM}(g_{\mu\nu},\Psi) + \mathcal L_\textrm{int} (g_{\mu\nu},\varphi,\Psi)\right]\, . \nonumber
\end{align}
with $\kappa=8\pi G/c^4$ where $G$ is Newton's constant, $R$ the curvature scalar of the space-time metric $g_{\mu\nu}$, $\varphi$ a dimensionless scalar field~\footnote{The dimensionless scalar field $\varphi$ is related to the scalar field $\phi$ of \cite{arvanitaki:2015qy,van-tilburg:2015fj} through $\varphi=\sqrt{4\pi G/c\hbar}~\phi=\sqrt{4\pi}\phi/M_\textrm{Pl}$, with $M_\textrm{Pl}$ the Planck mass, see also Eq.~(5) of~\cite{damour:2010zr}.}, $\mathcal L_\textrm{SM}$ is the Lagrangian density of the Standard Model of particles depending on the matter fields $\Psi$ and $\mathcal L_\textrm{int}$ parametrizes the interaction between the scalar field and matter. We will consider a quadratic scalar self-interaction
\begin{equation}\label{eq:V}
	V(\varphi)=2 \frac{c^2}{\hbar^2} m_\varphi^2 \varphi^2\, ,
\end{equation}
where the normalization of the potential has been chosen such that $m_\varphi$ has the dimension of a mass.

We consider  linear couplings between the scalar field and the matter fields similar to the ones introduced by Damour and Donoghue~\cite{damour:2010zr,damour:2010ve}~\footnote{We also provide general results (see Fig. \ref{fig:power}) that allow an easy evaluation of limits in other models e.g. with quadratic couplings \cite{Stadnik2015a}.}. The interacting part of the Lagrangian $\mathcal L_\textrm{int}$ is given by Eq.~(12) of~\cite{damour:2010zr}
\begin{align}
\mathcal L_\textrm{int} &=    \varphi\Big[\frac{d_e}{4\mu_0}F^2-\frac{d_g\beta_g}{2g_3}\left(F^A\right)^2 \\
&\qquad-c^2\sum_{i=e,u,d}(d_{m_i}+\gamma_{m_i}d_g)m_i\bar\psi_i\psi_i\Big]\nonumber \, ,
\end{align}
with $F_{\mu\nu}$ the standard electromagnetic Faraday tensor, $\mu_0$ the magnetic permeability, $F_{\mu\nu}^A$ the gauge invariant gluon strength tensor,  $g_3$ is the QCD gauge coupling, $\beta_3$ denotes the $\beta$ function for the running of $g_3$, $m_i$ the mass of the fermions, $\gamma_{m_i}$ the anomalous dimension giving the energy running of the masses of the QCD-coupled fermions and $\psi_i$ the fermions spinor. This Lagrangian is parametrized by five dimensionless coefficients $d_e, d_{m_e}, d_{m_u}, d_{m_d}$ and $d_g$ that characterize the coupling between the scalar and standard model fields. It is well-known that such a model will induce a violation of the Einstein Equivalence Principle for baryonic matter. This implies a violation of the WEP~\cite{damour:1994fk,damour:2010zr,damour:2010ve,damour:2012zr} as well as a violation of LPI through a modification of the gravitational redshift~\cite{damour:1997qv,*damour:1999fk,nordtvedt:2002uq,hees:2015ve,*Minazzoli:2015aa} and a space-time variation of the constants of Nature~\cite{damour:1994fk,nordtvedt:2002uq,bekenstein:1982zr,sandvik:2002ly,uzan:2011vn,olive:2008fk,minazzoli:2014xz,*hees:2014uq}. In particular, Damour and Donoghue~\cite{damour:2010zr,damour:2010ve} have shown that the particular form of the interacting Lagrangian leads to a linear dependence of 5 constants of Nature with respect to the scalar field
\begin{subequations}\label{eq:const_phi}
	\begin{align}
		\alpha(\varphi)&=\alpha(1+d_e\varphi) \, ,\\
		m_i(\varphi)&=m_i(1+d_{m_i}\varphi) \qquad \textrm{for } i=u,d,e\, , \\
		\Lambda_3(\varphi)&=\Lambda_3(1+d_g\varphi) \, , 
	\end{align}
where $\alpha$ is the fine structure constant, $m_i$ are the fermion (electron, up/down quark) masses and $\Lambda_3$ is the QCD mass scale. Note that the mean quark mass $m_q=(m_u+m_d)/2$ depends also linearly on the scalar field through~\cite{damour:2010zr,damour:2010ve}
\begin{equation*}
	m_q(\varphi)=m_q(1+d_{\hat m}\varphi)\, , \textrm{ with } d_{\hat m}=\frac{d_{m_u}m_u+d_{m_d}m_d}{m_u+m_d}\, .
\end{equation*}
\end{subequations}

The Klein-Gordon equation deriving from the action~(\ref{eq:action}) in a flat Friedmann-Lema\^itre-Robertson-Walker space-time is given by~\cite{damour:1994fk}
\begin{equation}
	\ddot \varphi+3H \dot \varphi + \frac{m_\varphi^2 c^4}{\hbar^2} \varphi=\frac{4\pi G}{c^2}\sigma \, ,
\label{KG}
\end{equation}
where $H$ is the Hubble constant and the dot denotes the derivative with respect to the cosmic time $t$.  The Hubble damping (due to the second term of (\ref{KG})) can safely be neglected as long as $m_\varphi >> \hbar H/c^2 \sim 1.5\times 10^{-33}$eV/c$^2$, and for experimental durations $<< 1/H$, with both conditions largely satisfied in our case. The source term $\sigma=\delta \mathcal L_\textrm{int}/\delta\varphi$ in Eq.~(\ref{KG}) is due to the non-minimal coupling between the scalar field and standard matter and is directly related to the baryonic matter density (see ~\cite{damour:1994fk}). Therefore, it will evolve with a characteristic time scale of $1/H$ and for periods much shorter, it can be considered as constant. Under these assumptions, the scalar field evolution is periodic
\begin{equation}\label{eq:phi_sol}
	\varphi = \frac{4\pi G \sigma \hbar^2 }{m_\varphi^2 c^6}+ \varphi_0 \cos (\omega t + \delta) \, , \textrm{ with } \omega=\frac{m_\varphi c^2}{\hbar}\, .
\end{equation}  
The oscillating part of the solution has been developed in \cite{arvanitaki:2015qy,van-tilburg:2015fj} where the source term has not been considered. The scalar field gives rise to a cosmological density $\rho_\varphi=\frac{c^2}{8\pi G}(\dot{\varphi}^2 + \frac{V(\varphi) c^2}{2})$ and a pressure $p_\varphi=\frac{c^2}{8\pi G}(\dot{\varphi}^2 - \frac{V(\varphi) c^2}{2})$. Substituting from (\ref{eq:phi_sol}) and (\ref{eq:V}) and averaging over one period of the cosine we find that the second term of (\ref{eq:phi_sol}) does not contribute to the pressure. It thus acts as a pressureless fluid which we identify as DM with density
\begin{equation}\label{eq:rho}
	\rho_{\tilde \varphi}=\frac{c^2}{4\pi G} \frac{\omega^2 \varphi_0^2}{2}=\frac{c^6}{4\pi G \hbar^2} \frac{m_\varphi^2 \varphi_0^2}{2} \, .
\end{equation}

\par\par Cosmological considerations place a lower limit of DM mass at $10^{-24}$ eV \cite{marsh:2015aa}. In addition, the scalar field oscillations have a finite coherence time given by $\tau_\textrm{coh}\sim 2\pi/\omega /(v/c)^2$ where $v/c\sim 10^{-3}$ (see also \cite{van-tilburg:2015fj}). In this analysis, the highest angular frequency considered is $3.6 \times 10^{-3} $ rad/s, which corresponds to a coherence time of $\tau_\textrm{coh}\sim 55$ years, much larger than the time span of our data.

\par As mentioned in \cite{arvanitaki:2015qy,derevianko:2016yu} and as can be seen directly from Eqs.~(\ref{eq:const_phi}), the scalar field oscillations from Eq.~(\ref{eq:phi_sol}) will produce similar oscillations on the fine structure constant, on the masses of the fermions and on the QCD mass scale. 

Atomic transition frequencies are sensitive to possible variations of the constants of the Standard Model. The variation of the frequency ratio X of two atomic transitions is characterized by $d\ln X=k_\alpha d\ln \alpha+k_\mu d \ln (m_e/m_p) + k_q d\ln (m_q/\Lambda_3)$ where the $k$'s represent sensitivity coefficients \cite{flambaum:2004fk}. Recent atomic structure calculations have shown that for the Rb/Cs ground state hyperfine transitions $k_\alpha=-0.49$, $k_\mu=0$ and $k_q=-0.021$ \cite{prestage:1995pl,flambaum:2006pr,dzuba:2008uq,dinh:2009fk}. In contrast, ratios of electronic-dipole transition frequencies, e.g. in optical clocks or in Dy, have only $k_\alpha \neq 0$ and are thus insensitive to variations of the other fundamental constants. The dependence of the Rb/Cs frequency ratio on $k_\alpha$ and $k_q$ associated with the harmonic evolution of the constants of Nature from Eqs.~(\ref{eq:const_phi}) and~(\ref{eq:phi_sol}) shows that the ratio of Rb/Cs hyperfine frequencies also exhibits a harmonic signature $y_{Rb}/y_{Cs}-1\approx\mathcal O+\mathcal C_\omega \cos \omega t+\mathcal S_\omega \sin\omega t =  \mathcal O+\mathcal A \cos (\omega t+\delta)$ where $y = \nu/\nu_0$ are the frequencies normalized to their nominal values and $\mathcal O$ is a constant offset. The amplitude of oscillation is given by
\begin{align}
\mathcal A&=\sqrt{\mathcal C_\omega^2+\mathcal S_\omega^2}= \left[k_\alpha d_e + k_q(d_{\hat{m}}-d_g)\right]\varphi_0\nonumber\\
	&=\left[k_\alpha d_e + k_q(d_{\hat{m}}-d_g)\right]\frac{1}{\omega}\left(\frac{8\pi G}{c^2}\rho_\textrm{DM}\right)^{1/2}  \, , \label{eq:amp}	
\end{align}
with $\rho_\textrm{DM}$ the DM energy density (in our galaxy, $\rho_\textrm{DM}\approx 0.4$ GeV/cm$^3$~\cite{mcmillan:2011vn}).	 In the last equation, we have assumed that the DM energy density is entirely due to the scalar field (see Eq.~(\ref{eq:rho})).

We use the dual $^{133}$Cs/$^{87}$Rb atomic fountain clock FO2 at LNE-SYRTE that operates simultaneously on both species thereby providing primary (Cs) and secondary (Rb) realizations of the SI second in parallel \cite{Guena2010,Guena2012,Guena2014}. A detailed description of the experimental apparatus can be found in \cite{Guena2010,Guena2012,guena:2012ys,Guena2014}, here we only recall the main features. Rb and Cs atoms are simultaneously laser cooled, launched, state selected, and probed with the Ramsey interrogation method, and finally selectively detected using time resolved laser-induced fluorescence, in the same vacuum chamber (see e.g. Fig.~2 of \cite{Guena2012}). The $|F=1,m_F=0\rangle \to |F=2,m_F=0\rangle$ hyperfine transition frequency of $^{87}$Rb at $\approx$6.8~GHz and the $|F=4,m_F=0\rangle \to |F=3,m_F=0\rangle$ hyperfine transition frequency of $^{133}$Cs at $\approx$9.2~GHz are simultaneously measured against the same ultrastable microwave reference at the 1.6 s fountain cycle, corrected for all known systematic effects (cold collisions, 2nd order Zeeman shifts, Blackbody radiation, etc... \cite{Guena2012,Guena2014}), and then averaged over synchronous intervals of $\Delta t_0 = 864$~s duration. 

\begin{figure}[htb]
\centering
 \includegraphics[width=0.45\textwidth]{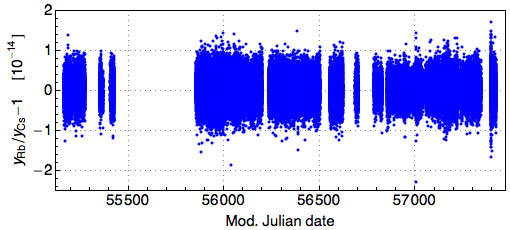}
\caption{Measurements of the normalized ratio of Rb/Cs frequencies at 864 s interval, spanning Nov. 2009 to Feb. 2016. Total of 100 	814 points with mean = $1.1\times 10^{-16}$ and standard deviation = $3\times 10^{-15}$. }
\label{fig:raw}
\end{figure}

\begin{figure}[htb]
    \centering
\includegraphics[width=0.45\textwidth]{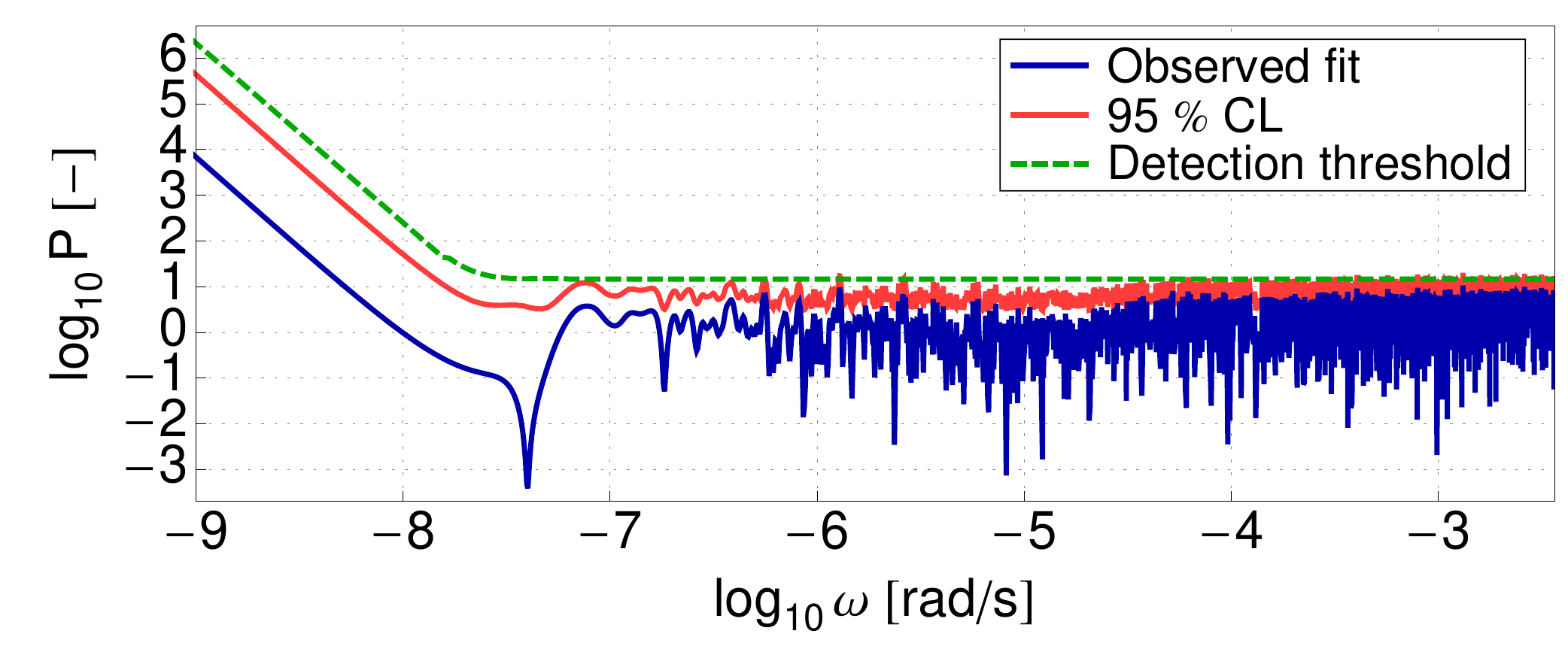}\\
\includegraphics[width=0.45\textwidth]{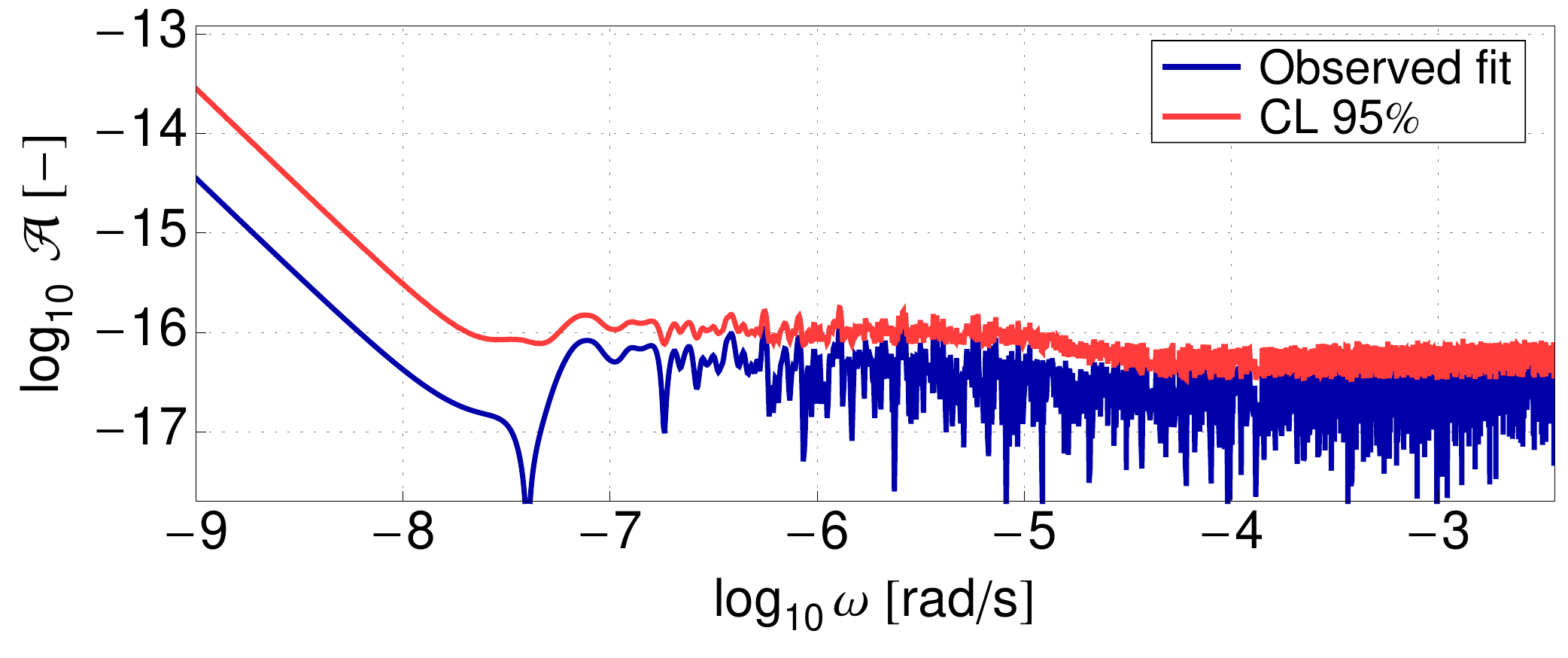} \\
\caption{Top: Normalized power spectrum (blue) obtained from the best-fit (see Eq.~(\ref{eq:power})) with the corresponding 5\% detection threshold (see text). The red line corresponds to the maximum allowed signal at 95\% confidence. \\
Bottom: Non-normalized amplitude spectrum $\mathcal A$ (blue) and corresponding maximum allowed signal at 95\% confidence (red).}
\label{fig:power}
\end{figure}

Our data set consists of measurements of $y_{Rb}/y_{Cs}$ spanning November 2009 to February 2016. The measurements are continuous with some gaps due to maintenance and investigation of systematics, giving an overall duty cycle of $\approx$~45\% over more than six years (see Fig.~\ref{fig:raw}).

The noise is roughly stationary over the complete data set, and characterized by white frequency noise with two different amplitudes depending on the averaging time (see Fig.~7 and related discussion in \cite{Guena2014}). This behavior is well understood and reproducible. It results from the operation of FO2 with atom numbers that are intentionally varied in order to correct for the collisional frequency shifts \cite{Guena2014}. The variance of our data $\sigma_o^2(\omega)$ depends on the Fourier frequency and is given by 
\begin{eqnarray}\label{eq:noise}
\sigma_o^2(\omega) &=& 4.6\times 10^{-29} \, , \textrm{ for } \omega\leq 9.0\times10^{-6} \textrm{ rad/s } \nonumber \\
\sigma_o^2(\omega) &=& 9.3\times 10^{-30} \, , \textrm{ for } \omega\geq 4.5\times10^{-5} \textrm{ rad/s } \nonumber \\
\sigma_o^2(\omega) &=& 4.2\times 10^{-34}/\omega \, , \textrm{ otherwise, } 
\end{eqnarray}
which is equivalent to the noise levels shown in \cite{Guena2014}.  

Our goal is to search for a sinusoidal signature in the $^{87}$Rb/$^{133}$Cs atomic frequency ratio measurements. Our methodology is similar to the one used in \cite{van-tilburg:2015fj}, is fully described in \cite{scargle:1982fk} and is presented in details in the supplemental material associated with this paper.  The highest analyzed angular frequency $\om$ is chosen to be $\pi/\Delta t_0\approx 3.6\times 10^{-3}$~rad/s. We can estimate the normalized power spectrum for each frequency
\begin{equation}\label{eq:power}
	P(\omega)=\frac{N_o}{4\sigma_o^2(\omega)}(\mathcal C_\omega^2+\mathcal S_\omega^2)\, ,
\end{equation}
where $N_o$ is the number of measurements and $\sigma_o^2(\omega)$ is their estimated variance given in (\ref{eq:noise}). In addition, a detection threshold has been estimated. This threshold is defined as the ensemble of power levels (for each frequency) such that the statistical probability of finding at least one power larger than that level in case of only noise is smaller than $p_0=5\%$, i.e. if at any frequency we find a value of the power spectrum larger than this threshold value and interpret it as a detection, the probability of it being a false detection is less than 5\%. 

In the top of Fig.~\ref{fig:power}, we present the results of this analysis for the Rb/Cs data set. Since the measured power spectrum is always smaller than the corresponding detection threshold, we conclude that there is no evidence of a harmonic modulation in our data. In the bottom of Fig.~\ref{fig:power}, we present the same results in terms of the amplitude of a hypothetical harmonic oscillation  $\mathcal A$ instead of the power spectrum. The figure shows the observed upper limit on the amplitude of a harmonic modulation allowed by the observations. These results can be directly used to constrain any model that predicts a periodic variation of the ratio of the Rb/Cs hyperfine frequencies, e.g. massive scalar fields with quadratic coupling to standard matter \cite{Stadnik2015a} (see also \cite{Stadnik:2016aa}). A detailed evaluation of all systematic effects that could affect the measured transition frequencies can be found in \cite{Guena2010,Guena2012,Guena2014}. A discussion specific to our search is presented in the supplemental material, the conclusion being that our results are limited predominantly by statistics rather than systematic effects.

\begin{figure}[htb]
\centering
\includegraphics[width=0.45\textwidth]{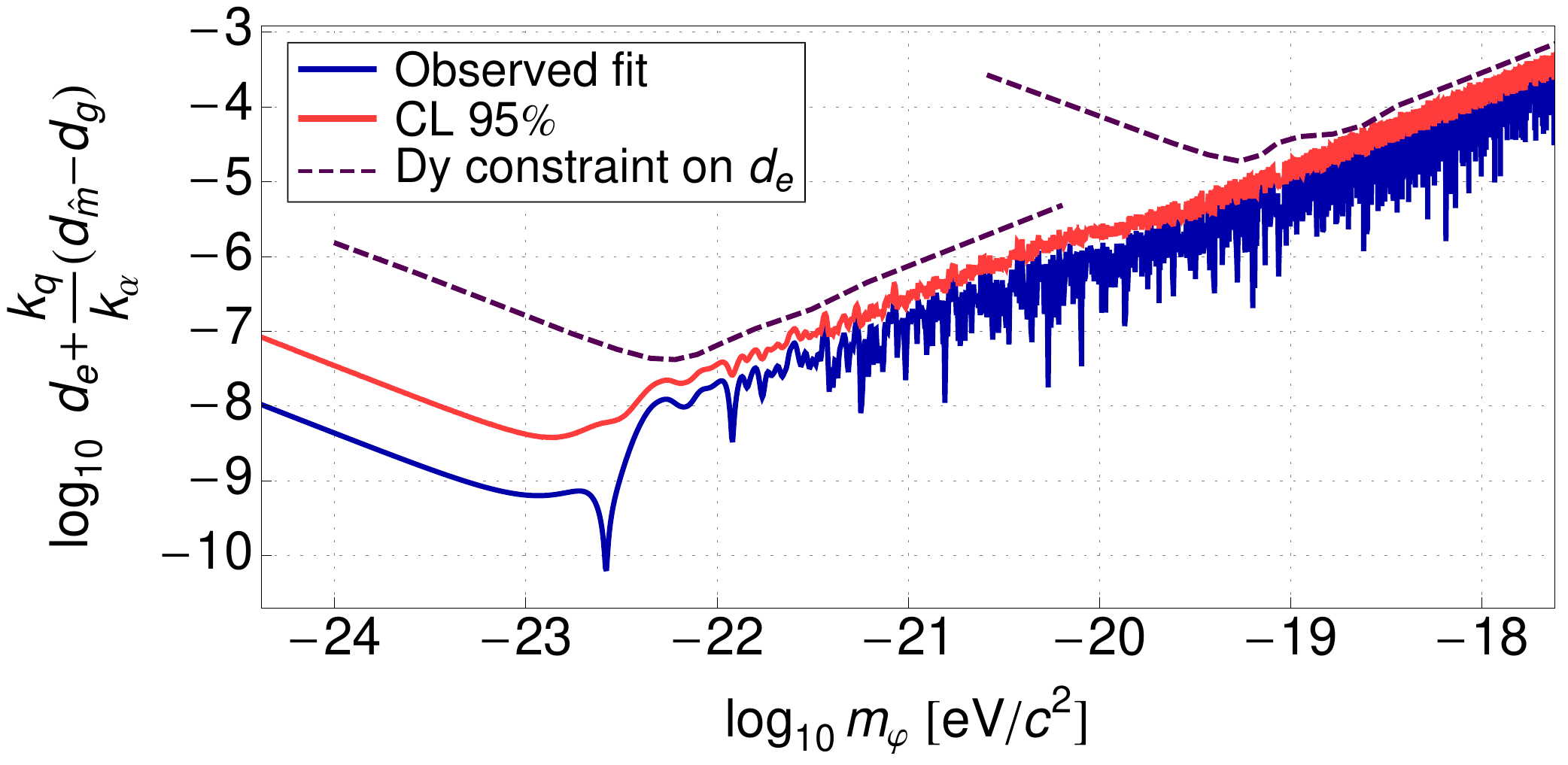}
\caption{Estimated values of the linear combination $d_e+k_q/k_\alpha(d_{\hat m}-d_g)=d_e+0.043(d_{\hat m}-d_g)$ of coupling constants $d_i$ between a massive scalar field and standard matter fields as a function of scalar field mass. The best fit values are shown in blue, with the 95\% confidence upper bounds in red. The purple dashed line represents the 95\% confidence upper bound obtained with Dy atoms in \cite{van-tilburg:2015fj}, which is only sensitive to $d_e$.}
\label{fig:d}
\end{figure}

Using Eq.~(\ref{eq:amp}), we can now transform our amplitude spectrum into limits on $d_e+k_q/k_\alpha(d_{\hat m}-d_g)=d_e+0.043(d_{\hat m}-d_g)$. Fig.~\ref{fig:d} shows our estimation and 95\% CL upper bound on this combination  as a function of the scalar field mass $m_\varphi=\hbar \omega/c^2$. We can exclude couplings larger than 5.3 $\times 10^{-4}$ at any $m_\varphi$ within our range, with our most stringent limit being as low as 3.8 $\times 10^{-9}$ at $m_\varphi=1.4 \times 10^{-23}$ eV/c$^2$. Our limits are complementary to those of \cite{van-tilburg:2015fj} and also to those coming from tests of the weak equivalence principle \cite{damour:2010zr} as they probe different combinations of the coupling constants $d_i$. If we assume that the scalar field is coupled only to electromagnetism (only $d_e \neq 0$) then our limits improve those of \cite{van-tilburg:2015fj} by more than an order of magnitude, and are far more stringent than those from WEP tests in the range of $m_\varphi$ considered here (which are of order of $10^{-3}$ \cite{arvanitaki:2016qv}).

\par In conclusion, massive scalar fields are a possible candidate for dark matter, and can be searched for by searching for a harmonic oscillation of fundamental constants, which in turn leads to an oscillation of frequency ratios of atomic transitions. In this letter, we have presented such a search, using over six years of precision measurements of the $^{87}$Rb/$^{133}$Cs ground state hyperfine frequency ratio at LNE-SYRTE. We see no evidence for an oscillating massive scalar field, but set upper limits on a linear combination of coupling constants between such a field and standard matter. Our results are complementary to previous measurements which constrain other parameter combinations, and improve previous results by over an order of magnitude when allowing only coupling to electromagnetism. The rapid progress of atomic clocks over the last years will allow similar searches with other types of transitions. That will further limit the parameter space for massive scalar fields as dark matter candidates and their coupling to standard matter.

\bibliography{dilaton}

\appendix
\section{Supplemental material}

\paragraph{Data analysis}
As mentioned in the main part of the paper, the noise in the measured $^{87}$Rb/$^{133}$Cs frequency ratio is roughly stationary over the complete data set, and characterized by white frequency noise with two different amplitudes depending on the averaging time (see Fig.~7 and related discussion in \cite{Guena2014}). This behavior is well understood and reproducible. It results from the operation of FO2 with atom numbers that are intentionally varied in order to correct for the collisional frequency shifts \cite{Guena2014}. The variance of our data $\sigma_o^2(\omega)$ depends on the Fourier frequency and is given by 
\begin{eqnarray}\label{eq:noiseS}
\sigma_o^2(\omega) &=& 4.6\times 10^{-29} \, , \textrm{ for } \omega\leq 9.0\times10^{-6} \textrm{ rad/s } \nonumber \\
\sigma_o^2(\omega) &=& 9.3\times 10^{-30} \, , \textrm{ for } \omega\geq 4.5\times10^{-5} \textrm{ rad/s } \nonumber \\
\sigma_o^2(\omega) &=& 4.2\times 10^{-34}/\omega \, , \textrm{ otherwise, } 
\end{eqnarray}
which is equivalent to the noise levels shown in \cite{Guena2014}.  

Our goal is to search for a sinusoidal signature in the Rb/Cs atomic frequency ratio measurements. Our methodology is similar to the one used in \cite{van-tilburg:2015fj} and is fully described in \cite{scargle:1982fk}. For each frequency $\omega$ we fit a signal of type $R(t)=\mathcal O_\omega+\mathcal C_\omega \cos \left(\omega t  \right)+\mathcal S_\omega \sin \left(\omega t  \right)$ to the sequence of measurements $R_i = y_{Rb}(t_i)/y_{Cs}(t_i)-1$. We perform a Bayesian inference of the parameters $\mathcal O_\omega, \mathcal C_\omega$ and $\mathcal S_\omega$ using a Monte Carlo (MC) algorithm \cite{gregory:2010qv}. We assume the observations to be independent and the errors to be normally distributed and we use flat prior probability distribution functions of the parameters.  The highest analyzed angular frequency $\om$ is chosen to be $\pi/\Delta t_0$. The Bayesian MC inference allows us to also analyze observations for angular frequencies smaller than $2\pi/T$  (with $T$ the total time span of the measurements) where the correlations between the parameters become significant. For $\omega > 2\pi/T$ we have verified independently that an ordinary least squares fit of $\mathcal O_\omega, \mathcal C_\omega$ and $\mathcal S_\omega$ yields the same result as the Bayesian MC algorithm.

For each occurrence of our MC sampling, we can estimate the normalized power spectrum
\begin{equation}\label{eq:powerS}
	P(\omega)=\frac{N_o}{4\sigma_o^2(\omega)}(\mathcal C_\omega^2+\mathcal S_\omega^2)\, ,
\end{equation}
where $N_o$ is the number of measurements and $\sigma_o^2(\omega)$ is their estimated variance given in (\ref{eq:noiseS}).

In addition, for each frequency, we perform a Bayesian MC inference of the parameters $\mathcal O_\omega, \mathcal C_\omega$ and $\mathcal S_\omega$ with $M(t_i)=0$, which we transform into a normalized power $P_e(\omega)$ using Eq.~(\ref{eq:powerS}). The obtained probability distribution of $P_e(\omega)$ is representative of the distribution in the case of white noise. For high frequencies $\omega > 2\pi/T$, the resulting noise power spectrum is equivalent to the modified periodogram introduced in \cite{scargle:1982fk} which is exponentially distributed. This allows a consistency check of our method by comparing the distribution obtained with our MC sampler with the theoretical one which is given by the cumulative distribution function Prob~$\big\{ P_e(\omega)<P_0\big\}=1-e^{-P_0}$. For low frequencies, $P_e(\omega)$ does not coincide with the modified periodogram of \cite{scargle:1982fk} anymore and its statistical properties are more complex (see also~\cite{van-tilburg:2015fj}). Therefore, we rely only on our Bayesian MC inference for our estimate at low frequencies ($\omega \leq 2\pi/T$).

For each frequency, we determine a detection threshold, which is the ensemble of power levels $P_{th}(\omega)$ such that the statistical probability of finding at least one power larger than that level in case of only noise is smaller than $p_0=5\%$, i.e. if at any frequency we find a value of $P(\omega)\geq P_{th}(\omega)$ and interpret it as a detection, the probability of it being a false detection is less than 5\%. We determine this level following Sec. III.c. of \cite{scargle:1982fk}. Considering a set of $N_\omega$ independent frequencies, the Bayesian inference allows us to determine the level $P_{th}(\omega)$ such that Prob~$\big\{P_e(\omega) < P_{th}(\omega) \big\}=(1-p_0)^{1/N_\omega}$. The probability Prob$_{th}$ that at least one value in the noise power spectrum is higher than its corresponding $P_{th}(\omega)$ is then given by 
\begin{equation}
	\textrm{Prob}_{th}=1-\prod_{j=1}^{N_\omega} 	\textrm{Prob}\big\{P_e(\omega_j)<P_{th}(\omega_j)\big\}=p_0 \, ,
\end{equation}
where the product is over the $N_\omega$ independent frequencies $\omega_j$. Once again, for high frequencies, we have checked that the obtained values correspond to the analytic results given in \cite{scargle:1982fk}.

Finally, we derive a 95\% confidence limit upper bound on the estimated power by taking the 95-th percentile value of our MC sampling of $P(\omega)$. This corresponds to the maximum allowed power (at 95\% confidence) given the actually measured power in our data and assuming that all of that is a signal (as also used in \cite{van-tilburg:2015fj}). We have checked that our estimated limit is consistent with the analytic expression from  \cite{scargle:1982fk} valid for high frequencies.

In the top of Fig.~2 of the main part of the paper, we present the results of this analysis for the Rb/Cs data set. Since the measured power spectrum is always smaller than the corresponding detection threshold, we can conclude that there is no evidence of a harmonic modulation. In the bottom of Fig.~2, we present the same results in terms of the amplitude of a hypothetical harmonic oscillation defined by $\mathcal A=\sqrt{\mathcal C_\omega^2+\mathcal S_\omega^2}$ instead of the power spectrum. The figure shows the observed upper limit on the amplitude of a harmonic modulation allowed by the observations.

\paragraph{Analysis of systematics}
Detailed evaluations of all systematic effects that could affect the measured transition frequencies are carried out regularly as described in detail in \cite{Guena2010,Guena2012,Guena2014}.  The impact of the scalar field on the corrected systematic effects can safely be neglected.  The reason is that most corrections (e.g. Blackbody radiation shift,  Doppler effects) do not involve measurements of the atomic frequency, they depend on e.g. temperature measurements which we do not expect to be affected by the DM scalar field. Those that do depend on atomic frequency for their evaluation involve a large ``leverage'' factor which suppresses the putative effect of the scalar field via the correction by many orders of magnitude with respect to the direct effect on the Rb/Cs frequency ratio. As an example, the 2nd order Zeeman effect (the clock transitions are insensitive to the 1st order Zeeman effect) is corrected by measuring the frequency of other transitions that are sensitive to 1st order effect to determine the magnetic field seen by the atoms \cite{Guena2010,Guena2012,Guena2014}. Those magnetic field measurements could be affected by a time varying scalar field and thus translate into a time varying correction that might mask the signal we are searching for. However, any such effect will be suppressed by a factor $2 B K_{Z2}/K_{Z1}$, where $B$ is the magnetic field, and $K_Z$ are the first and second order Zeeman coefficients. For the Cs hyperfine transition and our magnetic field of $\approx 200$~nT that factor is $\approx 2.4\times 10^{-6}$, with a similar value for Rb.

\begin{figure}[htb]
    \centering
\includegraphics[width=0.45\textwidth]{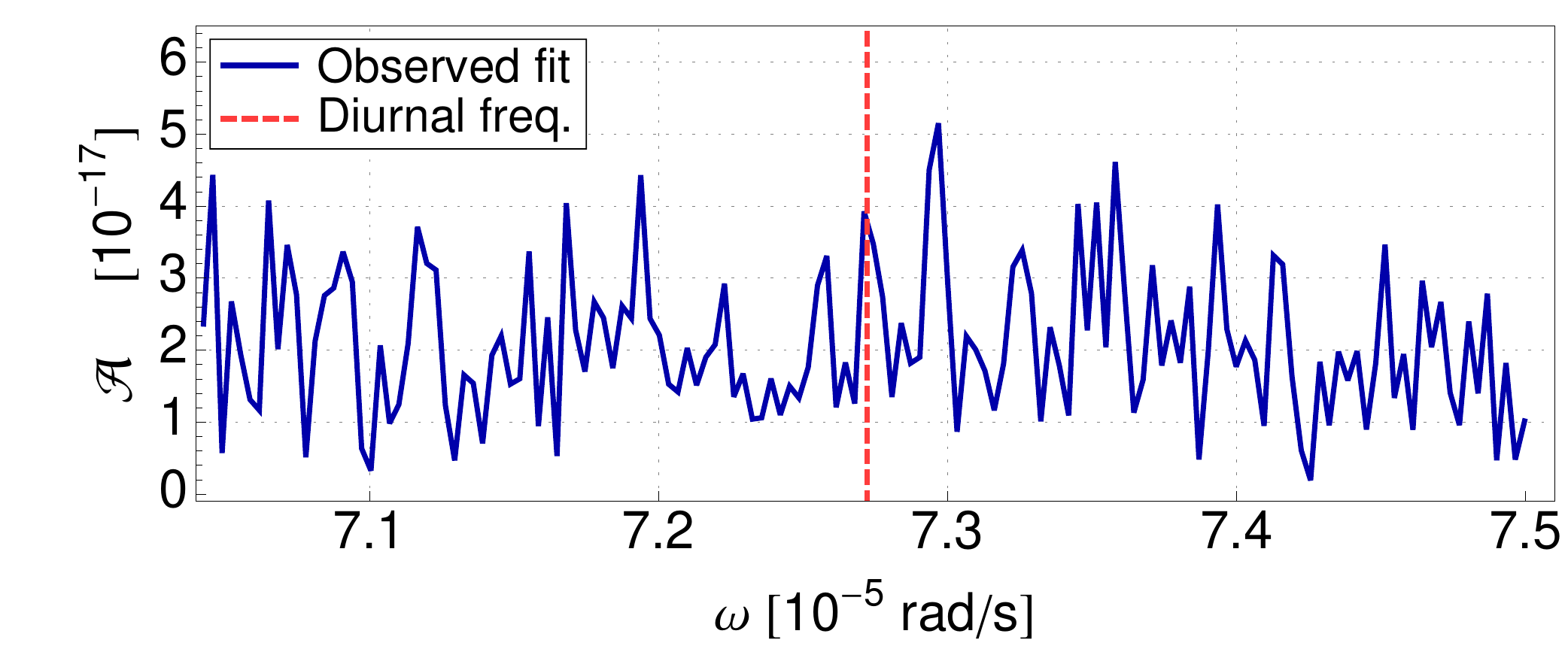}
\caption{Zoom around diurnal frequency for the estimated amplitude spectrum $\mathcal A$ from bottom of Fig. 2 of the main part of the paper.}
\label{fig:powerZ}
\end{figure}

From \cite{Guena2010,Guena2012,Guena2014}, the overall systematic uncertainties for the $^{133}$Cs and $^{87}$Rb transition are $u_\textrm{Cs}=2.1\times 10^{-16}$ and $u_\textrm{Rb}=3.2\times 10^{-16}$ in fractional frequency. The uncertainty on the difference is expected to be significantly lower given that some of the systematic effects are correlated and therefore partly cancel  (e.g. temperature, magnetic fields, \dots). Furthermore, we see no evidence of any systematic effect at the diurnal frequency where such effects are most likely to occur (see Fig.~\ref{fig:powerZ} from this supplemental material). We therefore conclude that our results are dominated by statistical uncertainties, as discussed above.

\end{document}